\documentclass[twocolumn,showpacs,amssymb,aps]{revtex4}
\usepackage{amsmath}
\usepackage{graphicx}
\begin{document}

\title{Hard thermal effective actions in the Schwinger formulation}  

\author{Ashok Das$^{a,b}$ and J. Frenkel$^{c}$}
\affiliation{$^{a}$ Department of Physics and Astronomy,
University of Rochester,
Rochester, NY 14627-0171, USA}
\affiliation{$^{b}$ Saha Institute of Nuclear Physics, 1/AF Bidhannagar, Calcutta 700064, INDIA}
\affiliation{$^{c}$ Instituto de F\'{\i}sica, Universidade de S\~ao
Paulo, S\~ao Paulo, SP 05315-970, BRAZIL}

\bigskip

\begin{abstract}

We derive the properties of hard thermal effective actions in gauge theories from the point of view of Schwinger's proper time formulation. This analysis is simplified by introducing a set of generalized energy and momenta which are conserved and are non-local in general. These constants of motion, which embody energy-momentum exchanges between the fields and the particles along their trajectories, can be related to a class of gauge invariant or covariant potentials in the hard thermal regime. We show that in this regime the generalized energy, which is non-local in general,  generates the characteristic non-local behavior of the hard thermal effective actions.
\end{abstract}


\maketitle

\section{Introduction}

The high temperature properties of the quark-gluon plasma in QCD are of great interest not only in their own right, but also as a starting point for the resummation of perturbation theory \cite{kapusta,lebellac}. These physical properties are encoded in the hard thermal effective action whose coefficient (the overall multiplicative factor) is proportional to $T^{2}$ where $T$ represents the equilibrium temperature. Such leading contributions to the effective action arise from one loop diagrams where the internal momentum is of order $T$ which is much larger than any external momentum. The corresponding hard thermal effective actions have been derived  from the point of view of thermal field theory \cite{braaten,frenkel,taylor,nair1,nair2,mckeon,mckeon2} as well as from the point of view of semi-classical transport equations \cite{elze,blaizot,nair,jackiw,kelley,pisarski5,litim}. It is known from these studies that the hard thermal effective actions are gauge invariant and, in general, are non-local except in the static limit where they become local.

The one-loop effective actions at zero temperature, on the other hand, are commonly derived using the proper time formulation of Schwinger \cite{schwinger} which is manifestly gauge invariant. However, this formulation is not as much developed at finite temperature \cite{jalilian}  and the main purpose of this short paper is to derive the hard thermal effective actions as well as  their properties from Schwinger's proper time approach. We find that, in this approach, all the information about the hard thermal effective action is contained in a set of conserved generalized momenta, which are in general non-local. In the hard thermal regime, these momenta can be related to a class of gauge invariant or covariant potentials \cite{gaugeinvariant}. In this regime, the generalized energy, which is in general non-local, contains all the information about the non-local behavior of the effective actions which become local only in the static limit. While our analysis holds for all theories, for brevity we discuss only gauge theories in this paper. In section {\bf II}, we recapitulate briefly Schwinger's proper time approach and derive the hard thermal effective action resulting from scalar QED. In this case, we construct the conserved generalized momenta and determine the set of associated gauge invariant potentials \cite{gaugeinvariant} in the hard thermal regime. In section {\bf III}, we extend this analysis to non-Abelian gauge theories and derive the corresponding generalized momenta and the related set of gauge covariant potentials in the hard thermal regime. These potentials, which generate correctly the relevant hard thermal contributions to all orders, are in general non-local. We conclude this note with a brief summary in section {\bf IV}.

\section{Effective action for QED}

Let us consider scalar QED in $3+1$ dimensions described by the Lagrangian density (our metric has the signature $(+,-,-,-)$)
\begin{equation}
{\cal L}  = \left((\partial_{\mu} - ie A_{\mu})\phi\right)^{\dagger} (\partial^{\mu} - ie A^{\mu})\phi - m^{2} \phi^{\dagger}\phi.\label{lagrangian}
\end{equation}
The Lagrangian density is quadratic in the scalar fields which can be integrated out in the path integral leading to the generating functional
\begin{eqnarray}
Z[A_{\mu}] & = & N \left[\det \left((\partial_{\mu} - ieA_{\mu})(\partial^{\mu} - ie A^{\mu}) + m^{2}\right)\right]^{-1} \nonumber\\
& = &\!\! \left[\det \frac{1}{2m}\!\!\left((\Pi_{\mu} - eA_{\mu})(\Pi^{\mu} - eA^{\mu}) - m^{2}\right)\!\right]^{-1}\!\!\!\!,
\end{eqnarray}
where we have identified the canonical momentum conjugate to the coordinate $x^{\mu}$ as $\Pi_{\mu} = -i\partial_{\mu}$ and have chosen a particular form for the normalization constant $N$. It follows now that the one-loop effective action for this theory can be written as
\begin{eqnarray}
\Gamma & = & i {\rm Tr}\ \ln\ \frac{1}{2m}\left((\Pi_{\mu}-eA_{\mu})(\Pi^{\mu} - e A^{\mu}) - m^{2}\right)\nonumber\\
 & = & i {\rm Tr}\ \ln H,\label{eff}
 \end{eqnarray}
where ``Tr" stands for trace over a complete set of states and we have identified
\begin{equation}
H = \frac{1}{2m} \left((\Pi_{\mu}-eA_{\mu})(\Pi^{\mu}-eA^{\mu}) - m^{2}\right).\label{h}
\end{equation}
The expression \eqref{eff} can lead to the effective action at zero as well as at finite temperature depending on the periodicity condition and the basis states chosen.

In Schwinger's approach, the effective action \eqref{eff} can be written (in a regularized manner) as
\begin{equation}
\Gamma = - i {\rm Tr} \int_{0}^{\infty} \frac{d\tau}{\tau}\ e^{-\tau H},\label{propertime}
\end{equation}
where $H$ (defined in \eqref{h}) can be thought of as the evolution operator for the proper time parameter $\tau$. Defining the kinematic momentum as
\begin{equation}
p_{\mu} = \Pi_{\mu} - eA_{\mu},\label{kinematic}
\end{equation}
we can determine the proper time evolution of the coordinates and momenta from the canonical commutation relations to be
\begin{eqnarray}
\frac{dx^{\mu}}{d\tau} & = & -i [x^{\mu}, H] = \frac{p^{\mu}}{m},\nonumber\\
\frac{dp_{\mu}}{d\tau} & = & -i [p_{\mu}, H] = \frac{e}{m} F_{\mu\nu} p^{\nu},\label{eqns}
\end{eqnarray}
where $F_{\mu\nu}$ denotes the Abelian field strength tensor. (It is worth pointing out here that we have not worried about the order of factors in the second equation, which is not relevant in the hard thermal regime that we are interested in.) If the dynamical equations \eqref{eqns} for such a particle can be solved in a closed form, one can construct a complete set of states and evaluate the effective action \eqref{propertime} in a closed form. In general, this is not possible for a particle interacting with an arbitrary external field and in such a case, one studies the effective action in a perturbative manner \cite{schwinger}. However, as we will show, in the hard thermal regime where $\Pi \gg eA$, there is a great simplification yielding the leading result for the effective action in a straight forward manner. 

To calculate the hard thermal effective action, we note that the current of the theory at zero temperature follows from \eqref{eff} to be
\begin{eqnarray}
j^{\mu (0)} (x)\!\! & = & \frac{\delta \Gamma^{(0)}}{\delta A_{\mu} (x)} \nonumber\\
 \!\!& = & \!\!-2e\!\! \int\!\! \frac{d^{4}\Pi}{(2\pi)^{4}} (\Pi^{\mu}-eA^{\mu}) \frac{i}{(\Pi-eA)^{2} - m^{2}}.\label{current}
\end{eqnarray}
If we know the current, it can, of course, be functionally integrated (in principle) to yield the effective action. We note that 
the denominator in \eqref{current} can be thought of as an effective scalar propagator in a space-time dependent background field $A_{\mu}(x)$. In order to calculate the hard thermal (retarded) effective action, we need to define the current at finite temperature by generalizing this propagator to the appropriate finite temperature one through a suitable analytic continuation \cite{kapusta,lebellac,das} (in either imaginary time or real time formalism). Because of the presence of the background field in the zero temperature propagator, such a generalization is not immediately obvious. However, the extension of the propagator to finite temperature can be carried out as follows. 

Let us note from \eqref{eqns} that when interactions are present, the momentum $p_{\mu}$ (or $\Pi_{\mu}$) is not conserved. However, even in such a case, we can define a generalized momentum that is a constant of  motion in the following way. We note from \eqref{eqns} that the time evolution of functions in the phase space is given by the operator
\begin{equation}
\frac{d}{d\tau} = \frac{1}{m}\left((p\cdot \partial) + e F_{\mu\nu} p^{\nu} \frac{\partial}{\partial p_{\mu}}\right).\label{evolution}
\end{equation}
Let us next define the (non-local) operator (we note here parenthetically that this is, in fact, the operator that arises in the conventional calculations of the hard thermal effective action for QED)
\begin{equation}
K = \frac{e}{p\cdot \partial}\  F_{\mu\nu} p^{\nu}\frac{\partial}{\partial p_{\mu}}.
\end{equation}
Then, it can be easily checked with the help of \eqref{evolution} that
\begin{eqnarray}
P_{\mu} & = & p_{\mu} + Y_{\mu} = p_{\mu} - \frac{1}{1+K}\ Kp_{\mu}\nonumber\\
 & = & \frac{1}{1+K}\ p_{\mu},\label{P}
 \end{eqnarray}
 is conserved, namely,
 \begin{equation}
 \frac{dP_{\mu}}{d\tau} = 0.
 \end{equation}
The coordinate, canonically conjugate to this generalized momentum, can also be derived and has the form
\begin{equation}
X^{\mu} = x^{\mu} + \frac{1}{1+K}\ \frac{1}{p\cdot \partial}\ Y^{\mu},\label{X}
\end{equation}
and it satisfies the equation
\begin{equation}
\frac{dX^{\mu}}{d\tau} = \frac{P^{\mu}}{m}.
\end{equation}
Thus, we see that in these generalized variables, the dynamical equations reduce to those of free particle motion.

We note that this generalized momentum, $P_{\mu}$, is in general non-local. However, it can be easily verified that
\begin{equation}
P_{\mu}P^{\mu} = p_{\mu}p^{\mu},\label{relation}
\end{equation}
so that the zero temperature propagator can be written in these variables as a free propagator
\begin{equation}
i\Delta^{(0)} = \frac{i}{(\Pi - eA)^{2} - m^{2}} = \frac{i}{P^{2} - m^{2}},
\end{equation}
where we have used both \eqref{kinematic} as well as \eqref{relation}. The generalization to finite temperature is now immediate
\begin{equation}
i\Delta^{(T)} = \frac{i}{P^{2} - m^{2}} + 2\pi n(|P_{0}|) \delta (P^{2}-m^{2}),\label{T}
\end{equation}
where $n (|P_{0}|)$ denotes the bosonic distribution function. We note here in passing that since the thermal propagator is related to the zero temperature one through the thermal operator \cite{silvana}, the currents at finite temperature will also satisfy such a relation.

Using \eqref{T} we can determine the temperature dependent part of the current (see \eqref{current}) (we are using the notation $j^{\mu (T)} = j^{\mu (0)} + j^{\mu (\beta)}$)
\begin{widetext}
\begin{eqnarray}
j^{\mu (\beta)} & = & -2e \int \frac{d^{4}\Pi}{(2\pi)^{4}} (\Pi^{\mu} - eA^{\mu}) 2\pi n(|P_{0}|) \delta (P^{2}-m^{2})\nonumber\\
& = & -2e \int \frac{d^{4}P}{(2\pi)^{4}} \left(1 - \frac{e}{(P\cdot\partial)^{2}} P_{\sigma}\partial_{\rho}F^{\rho\sigma} + \cdots\right)\left(P^{\mu} + \frac{e}{P\cdot\partial} P_{\alpha}F^{\mu\alpha}+\cdots\right) 2\pi n(|P_{0}|) \delta (P^{2}-m^{2}),
\end{eqnarray}
\end{widetext}
where the first factor arises from the Jacobian for the change of variables. In the hard thermal limit where we can neglect masses and assume $P\gg \partial$, the leading contribution to the current occurs only at order $e^{2}$ and takes the form
\begin{eqnarray}
j^{\mu (\beta)}_{HTL} &=& -2e^{2} \int \frac{d^{4}P}{(2\pi)^{3}} \delta (P^{2}) n(|P_{0}|)\nonumber\\
 &  & \quad \times \left(\eta^{\mu\nu} - \frac{P^{\mu}\partial^{\nu}}{P\cdot \partial}\right) \frac{P^{\rho}}{P\cdot\partial} F_{\nu\rho}.\label{HTL}
\end{eqnarray}
Using the standard integral
\begin{equation}
\int_{0}^{\infty} dx\ x n(x) = \frac{\pi^{2}T^{2}}{6},
\end{equation}
and integrating the current \eqref{HTL}, we obtain the temperature dependent hard thermal effective action to be
\begin{equation}
\Gamma_{HTL}^{(\beta)} =  \frac{e^{2}T^{2}}{12} \int d^{4}x \int \frac{d\Omega}{4\pi} F^{\mu\nu} \frac{\hat{p}_{\nu}\hat{p}^{\sigma}}{(\hat{p}\cdot \partial)^{2}} F_{\sigma\mu},\label{HTLaction}
\end{equation}
where we have labeled the variable of integration in \eqref{HTL} as $p$ and have defined 
\begin{equation}
\hat{p}^{\mu} = (1, \hat{\mathbf{p}}),\label{unitvector}
\end{equation} 
and $\int d\Omega$ denotes the angular integration over the unit vector $\hat{\mathbf{p}}$.

The integrand of the effective action in \eqref{HTLaction} is manifestly gauge invariant and Lorentz invariant, but appears to be manifestly non-local as well. The locality/non-locality of the hard thermal effective action can be best understood in terms of the kinematic momentum. In this variable, the hard thermal current \eqref{HTL} takes the form
\begin{equation}
j^{\mu (\beta)}_{HTL} = -2e \int \frac{d^{4}p}{(2\pi)^{3}} p^{\mu} \delta(p^{2}) n(|P_{0}|),\label{alternate}
\end{equation}
and we see that all the non-locality of the current (and, therefore, the effective action) is contained in the generalized energy $P_{0}$ and manifests through the dependence of the integrand on the distribution function $n(|P_{0}|)$. To understand the non-local behavior of $P_{0}$ better, let us write \eqref{P} as
\begin{equation}
P_{\mu} = p_{\mu} + e {\cal A}_{\mu},
\end{equation}
where in the hard thermal regime, we can identify
\begin{equation}
{\cal A}_{\mu} (x, \hat{p}) =  \frac{1}{\hat{p}\cdot\partial} \hat{p}^{\nu}F_{\nu\mu} = A_{\mu} - \frac{1}{\hat{p}\cdot\partial} \partial_{\mu} \hat{p}\cdot A.\label{potential}
\end{equation}
In this regime, we recognize ${\cal A}_{\mu}$ to correspond to the class of gauge invariant potentials \cite{gaugeinvariant} which are, in general, path dependent and non-local. For example, using the  retarded path integral form for the operator $\frac{1}{\hat{p}\cdot\partial}$ \cite{lifshitz}, we can explicitly express these potentials as 
\begin{eqnarray}
{\cal A}_{\mu} (x, \hat{p}) & = & - \int_{-\infty}^{t} dt'\ F_{\mu} (t', \mathbf{x} - \hat{\mathbf{p}} (t-t'))\nonumber\\
& = & - \int_{-\infty}^{t} dt'\ F_{\mu} (x'(t')),
\end{eqnarray}
where $eF_{\mu} = e \hat{p}^{\nu} F_{\mu\nu}$ denotes the Lorentz force four vector 
\begin{equation}
eF^{\mu} = e (\hat{\mathbf{p}}\cdot \mathbf{F}, \mathbf{F}),\quad \mathbf{F} = \mathbf{E} + \hat{\mathbf{p}}\times \mathbf{B},
\end{equation}
with $\mathbf{E},\mathbf{B}$ representing the electric and the magnetic fields respectively. It follows, therefore, that
\begin{eqnarray}
e{\cal A}_{0} (x, \hat{p})\!\! & = &\!\! - e \int_{-\infty}^{t} \!\!dt'\ \hat{\mathbf{p}}\cdot \mathbf{E} (x'(t')),\nonumber\\
e\boldsymbol{\cal A} (x, \hat{p})\!\! & = &\!\! - e\int_{-\infty}^{t} \!\!\!\!dt' \left(\mathbf{E} (x'(t')) + \hat{\mathbf{p}}\times \mathbf{B} (x'(t'))\!\right)\!,
\end{eqnarray}
can be interpreted in this case as the energy and momentum exchanged between the particle and the field along a trajectory parallel to $\mathbf{p}$. As is clear, in general the gauge invariant potentials are non-local. However, we note from \eqref{potential} that in the static limit (when the background field is static) ${\cal A}_{0}$ and, therefore, $P_{0}$ is local. It then follows from \eqref{alternate} that although the hard thermal current as well as the effective action are non-local in general,  they become local only in the static limit. It is interesting to remark here that integrating by parts \eqref{HTLaction} and using \eqref{potential}, the hard thermal effective action may be expressed in terms of the gauge invariant potentials  in the simple form
\begin{equation}
\Gamma_{HTL}^{(\beta)} =  \frac{m_{\rm ph}^{2}}{2} \int d^{4}x \int \frac{d\Omega}{4\pi}\ {\cal A}_{\mu} (x,\hat{p}) {\cal A}^{\mu} (x,\hat{p}),\label{schwinger}
\end{equation}
where $m_{\rm ph} = \frac{eT}{\sqrt{6}}$ represents the thermal mass of the photon. The form \eqref{schwinger} is reminiscent of the gauge invariant mass generated in the Schwinger model (at zero temperature) \cite{schwinger1}.

\section{Effective action for Yang-Mills theory}

The derivation of the hard thermal effective action (and its properties) for the Yang-Mills theory follows in a completely parallel manner to the discussion in the last section. Therefore, we only give a brief description of the essential steps involved in such a derivation. As in section {\bf II}, let us consider a complex scalar field in a given representation of $SU(N)$ interacting with a background non-Abelian gauge field. The Lagrangian density has the form (compare with \eqref{lagrangian})
\begin{equation}
{\cal L} = \left((\partial_{\mu} - ig A_{\mu}^{a}T^{a})\phi\right)^{\dagger} (\partial^{\mu} - ig A^{\mu b}T^{b})\phi - m^{2} \phi^{\dagger}\phi.
\end{equation}
Here the scalar field is a matrix of $SU(N)$ and $T^{a}, a=1,2,\cdots, N^{2}-1$ represent the generators of the group (in the representation to which the scalar fields belong) satisfying the commutation relations
\begin{equation}
[T^{a}, T^{b}] = i f^{abc} T^{c},\label{algebra}
\end{equation}
with $f^{abc}$ denoting the structure constants. Following the discussion in section {\bf II}, we can write the one-loop effective action as (see \eqref{eff})
\begin{equation}
\Gamma = i {\rm Tr}\ \ln H,\label{nonabelianeff}
\end{equation}
with the Hamiltonian given by
\begin{equation}
H = \frac{1}{2m} \left((\Pi_{\mu} - gA_{\mu}^{a}T^{a})(\Pi^{\mu} - g A^{\mu b}T^{b}) - m^{2}\right).\label{nonabelianh}
\end{equation}
We note that ``Tr" in \eqref{nonabelianeff} denotes summing over a complete set of states as well as a sum over the matrix indices of the group.

The effective action in \eqref{nonabelianeff} can now be given a proper time representation as in \eqref{propertime} and the dynamical proper time evolutions for the coordinates and momenta can be determined to be
\begin{eqnarray}
\frac{dx^{\mu}}{d\tau} & = & - i [x^{\mu}, H] = \frac{p^{\mu}}{m},\nonumber\\
\frac{dp_{\mu}}{d\tau} & = & - i [p_{\mu}, H] = \frac{g}{m} F_{\mu\nu}^{a}T^{a}p^{\nu},\label{nonabelianeqns}
\end{eqnarray}
where $F_{\mu\nu}^{a}$ denotes the non-Abelian field strength tensor
\begin{equation}
F_{\mu\nu}^{a} = \partial_{\mu}A_{\nu}^{a} - \partial_{\nu}A_{\mu}^{a} + g f^{abc} A_{\mu}^{b}A_{\nu}^{c},
\end{equation}
and we have identified the kinematic momentum to be
\begin{equation}
p_{\mu} = \Pi_{\mu} - gA_{\mu}^{a}T^{a}.
\end{equation}
We note that since the generators $T^{a}$ do not commute (see \eqref{algebra}), in this case, in addition to the usual equations of motion \eqref{nonabelianeqns} for the coordinates and momenta of the particle,  we will also have
\begin{equation}
\frac{dT^{a}}{d\tau} = -i [T^{a}, H] = -\frac{g}{m} f^{abc} (p\cdot A^{b}) T^{c}.\label{charge}
\end{equation}
Together with \eqref{charge}, therefore, the equations in \eqref{nonabelianeqns} describe a ``spinning" particle in the internal space and this ``spin" becomes an additional degree of freedom in this case. (We mention here again that we have disregarded the ordering of the factors which are not relevant in the hard thermal regime.) Furthermore, we remark here that the expectation value of the generators in the semiclassical limit (for large quantum numbers) can be identified with the color charge of the classical particle \cite{wong,brown}.

As in the last section, we note that neither the canonical momentum nor the kinematic momentum is conserved. However, we can determine a generalized momentum (see \eqref{P}) which is conserved in the following way. Let us define a derivative operator
\begin{equation}
\tilde{D}_{\mu} = \partial_{\mu} - g f^{abc} A_{\mu}^{b} T^{c} \frac{\partial}{\partial T^{a}}.
\end{equation}
It is easy to see that acting on the space of functions of the kind $f^{a} (x, p) T^{a}$, this gives
\begin{equation}
\tilde{D}_{\mu} \left(f^{a} (x,p) T^{a}\right) = \left(D_{\mu} f (x,p)\right)^{a} T^{a},\label{equivalence}
\end{equation}
where the covariant derivative is defined to be
\begin{equation}
\left(D_{\mu} f (x,p)\right)^{a} = \partial_{\mu} f^{a} + g f^{abc} A_{\mu}^{b} f^{c}. \label{covariant}
\end{equation}
With these and \eqref{nonabelianeqns} as well as \eqref{charge}, it can be shown that the conserved generalized momentum takes the form
\begin{eqnarray}
P_{\mu} & = & p_{\mu} + Y_{\mu} = p_{\mu} - \frac{1}{1+K} Kp_{\mu}\nonumber\\
 & = & \frac{1}{1+K}\ p_{\mu},\label{nonabelianP}
\end{eqnarray}
where in the present case
\begin{equation}
K = \frac{g}{p\cdot\tilde{D}} \left(F_{\mu\nu}^{a}T^{a}p^{\nu}\frac{\partial}{\partial p_{\mu}}\right).\label{K}
\end{equation}
Although these generalized conserved momenta are, in general, non-local, as in \eqref{relation} it can be verified that
\begin{equation}
P_{\mu}P^{\mu} = p_{\mu}p^{\mu}.
\end{equation}
Therefore, following the analysis of the previous section, we can write the temperature dependent part of the current in the hard thermal regime as (see \eqref{alternate})
\begin{equation}
j_{HTL}^{\mu a (\beta)} = -2g \int \frac{d^{4}p dT}{(2\pi)^{3}}\ T^{a} p^{\mu} \delta (p^{2}) n(|P_{0}|).\label{nonabelianHTL}
\end{equation}
Here $\int dT$ denotes the integration over the ``spin" degrees of freedom (color charge). Once again we emphasize here that the hard thermal current and, therefore, the action will be non-local in general simply because $P_{0}$ (defined in \eqref{nonabelianP}) has this behavior. 

To understand this non-local behavior better, we note that in the hard thermal regime, we can write (see \eqref{nonabelianP})
\begin{equation}
P_{\mu} = p_{\mu} + g{\cal A}_{\mu}^{a}T^{a},\label{np}
\end{equation}
where with the help of \eqref{equivalence} we find
\begin{equation}
{\cal A}_{\mu}^{a} (x, \hat{p}) = \left(\frac{1}{\hat{p}\cdot D} \hat{p}^{\nu}F_{\nu\mu}\right)^{a} = A_{\mu}^{a} - \left(\frac{1}{\hat{p}\cdot D} \partial_{\mu} \hat{p}\cdot A\right)^{a}.\label{nonabelianpotential}
\end{equation}
Here $D_{\mu}$ represents the covariant derivative defined in \eqref{covariant}. The gauge covariant potentials in \eqref{nonabelianpotential} represent a natural extension of the class of invariant potentials in \eqref{potential}. These non-Abelian potentials, which are in general non-local and path dependent, generate correctly the leading hard thermal contributions to all orders in $g$. Furthermore, as in the case of QED, we note that when the background field is static, the generalized energy in \eqref{np} has  the simple local form
\begin{equation}
P_{0} = p_{0} + g A_{0}^{a}T^{a},
\end{equation}
so that only in this limit, the hard thermal current as well as the effective action become local. In fact, in this limit only the gluon self-energy is leading at high temperature. In general, since $P_{0}$ is conserved, $n(|P_{0}|)$ is also constant under $\tau$ evolution. Using this fact, it can be readily shown that the hard thermal current \eqref{nonabelianHTL} is covariantly conserved,
\begin{equation}
D_{\mu}j_{HTL}^{\mu a (\beta)} = \partial_{\mu} j_{HTL}^{\mu a (\beta)} + g f^{abc} A_{\mu}^{b} j_{HTL}^{\mu c (\beta)} = 0.
\end{equation}
As a result, the hard thermal effective action, which is obtained by functionally integrating the hard thermal current is gauge invariant. The explicit form of the hard thermal effective action for the Yang-Mills fields can be written as \cite{braaten,frenkel}
\begin{equation}
\Gamma_{HTL}^{(\beta)} =  \frac{N(gT)^{2}}{12} \int d^{4}x \int \frac{d\Omega}{4\pi} \ F^{\mu\nu a} \left(\frac{\hat{p}_{\nu}\hat{p}^{\sigma}}{(\hat{p}\cdot D)^{2}}\right)^{ab} F_{\sigma\mu}^{b}.\label{HTLnonabelianaction}
\end{equation}
The similarity of the above action with \eqref{HTLaction} is worth noting. It is interesting to observe that integrating this expression by parts,  we can rewrite the hard thermal effective action \eqref{HTLnonabelianaction} in terms of the gauge covariant potentials \eqref{nonabelianpotential} in the simple form
\begin{equation}
\Gamma_{HTL}^{(\beta)} =  \frac{m_{\rm gl}^{2}}{2} \int d^{4}x \int \frac{d\Omega}{4\pi} \ {\cal A}_{\mu}^{a} (x,\hat{p}) {\cal A}^{\mu a} (x,\hat{p}),\label{simpleform}
\end{equation}
where  $m_{\rm gl} = \sqrt{\frac{N}{6}} gT$ is the thermal gluon mass, in complete analogy with \eqref{schwinger}. Alternatively, one could have started from a manifestly gauge invariant form like \eqref{simpleform}, which represents a natural generalisation of the QED action \eqref{schwinger}. Then, noticing that the overall multiplicative factor in \eqref{simpleform} can be uniquely determined by an explicit evaluation of the gluon self-energy, one would be readily led to the well known hard thermal effective action \eqref{HTLnonabelianaction}.

\section{Conclusion}

In this work, we have discussed the hard thermal effective actions and their properties from the point of view of Schwinger's proper time approach. We have shown that the non-localities of the current as well as the effective action can be understood from the behavior of a set of generalized momenta, $P_{\mu}$, which are conserved (even in the presence of interactions) and include the exchange of energy and momentum between the background fields and the particle along its trajectory. These conserved momenta are in general non-local, but satisfy the condition $P_{\mu}P^{\mu} = p_{\mu}p^{\mu}$, where $p_{\mu}$ denotes the kinematic momentum. As a result, the Lorentz invariant effective actions at zero temperature, which are functions of $P^{2}$, will be local as expected. On the other hand, at finite temperature, Lorentz invariance is broken because the rest frame of the heat bath defines a preferred reference frame. Consequently, the thermal effective actions will depend on the non-local energy $P_{0}$ through the distribution function. It is precisely this feature that generates the non-local behavior of the effective action in the hard thermal loop regime. An exception occurs only when the background is static, in which case the exchanged energy is local, and this is reflected in the local forms of the corresponding hard thermal effective actions. Although our discussion has been completely within the context of gauge field backgrounds, this analysis holds as well when the background involves scalar fields. The approach outlined in this paper may also be useful to study the properties of effective actions describing non-equilibrium systems at high temperature \cite{strickland}.

\vskip 1cm

\noindent{\bf Acknowledgment:}

This work was
supported in part by US DOE Grant number DE-FG 02-91ER40685, by CNPq
and FAPESP, Brazil.


\begin{thebibliography}{10}

\bibitem{kapusta} J. I. Kapusta, {\em Finite Temperature Field Theory} (Cambridge University Press, Cambridge, England, 1989).

\bibitem{lebellac} M. Le Bellac, {\em Thermal Field Theory} (Cambridge University Press, Cambridge, England, 1996).

\bibitem{braaten} E. Braaten and R. Pisarski, Nucl. Phys. {\bf B337}, 569 (1990); {\em ibid} {\bf B339}, 310 (1990); {\em ibid} Phys. Rev. {\bf D45}, 1827 (1992). 

\bibitem{frenkel} J. Frenkel and J. C. Taylor, Nucl. Phys. {\bf B334}, 199 (1990); {\em ibid} {\bf B374}, 156 (1992); {\em ibid} {\bf B685}, 393 (2004). 

\bibitem{taylor} J. C. Taylor and S. M. Wong, Nucl. Phys. {\bf B346}, 115 (1990).

\bibitem{nair1} E. Efraty and V. P. Nair, Phys. Rev. {\bf D47}, 5601 (1993).

\bibitem{nair2} R. Jackiw and V. P. Nair, Phys. Rev. {\bf D48}, 4991 (1993). 

\bibitem{mckeon} D. McKeon and A. Rebhan, Phys. Rev. {\bf D47}, 5487 (1993); {\em ibid} {\bf D49}, 1047 (1994).

\bibitem{mckeon2} F. T. Brandt and D. McKeon, Phys. Rev. {\bf D54}, 6435 (1996).

\bibitem{elze} H. Elze and U. Heinz, Phys. Rep. {\bf 183}, 81 (1989).

\bibitem{blaizot} J. P. Blaizot and E. Iancu, Nucl. Phys. {\bf B390}, 589 (1993); {\em ibid} {\bf B417}, 608 (1994); {\em ibid} {\bf B434}, 662 (1994); {\em ibid} Phys. Rep. {\bf 359}, 355 (2002).

\bibitem{nair} V. P. Nair, Phys. Rev. {\bf D48}, 3432 (1993); {\em ibid} {\bf D50}, 4201 (1994).

\bibitem{jackiw} R. Jackiw, Q. Liu and C. Lucchesi, Phys. Rev. {\bf D49}, 6787 (1994).

\bibitem{kelley} P. F. Kelly, Q. Liu, C. Lucchesi and C. Manuel, Phys. Rev. Lett. {\bf 72}, 3461 (1994); {\em ibid} Phys. Rev. {\bf D50}, 4209 (1994).

\bibitem{pisarski5} R. Pisarski, ''Non-abelian Debye screening, tsunami waves and worldline fermions,'' hep-ph/9710370. 

\bibitem{litim} D. Litim and C. Manuel, Phys. Rep. {\bf 364}, 451 (2002).

\bibitem{schwinger} J. Schwinger, Phys. Rev. {\bf 82}, 664 (1951).

\bibitem{jalilian} J. Jalilian-Marian, S. Jeon, R. Venugopalan and J. Wirstam, Phys. Rev. {\bf D62}, 045020 (2000). 

\bibitem{gaugeinvariant} P. A. M. Dirac, Can. J. Phys. {\bf 33}, 650 (1955); B. De Witt, Phys. Rev. {\bf 125}, 2189 (1962); S. Mandelstam, Ann. Phys. {\bf 19}, 1 (1962); A. Sisakyan, O. Shevchenko and I. Solovstov, Sov. J. Part. Nucl. {\bf 21}, 285 (1990); P. Gaete, Z. Phys. {\bf C76}, 355 (1997).

\bibitem{das} A. Das, {\em Finite Temperature Field Theory} (World Scientific, Singapore, 1997).

\bibitem{silvana} F. T. Brandt, A. Das, O. Espinosa, J. Frenkel and S. Perez, Phys. Rev. {\bf D72}, 085006 (2005); {\em ibid} {\bf D73}, 065010 (2006); {\em ibid} {\bf D73}, 067702 (2006).

\bibitem{lifshitz} E. Lifshitz and L. Pitaevski, {\em Physical Kinetics}, chapter 3 (Pergamon Press, Oxford, England, 1981).

\bibitem{schwinger1} J. Schwinger, Phys. Rev. {\bf 128}, 2425 (1962).

\bibitem{wong} S. K. Wong, Nuovo Cim. {\bf 65A}, 689 (1970).

\bibitem{brown} L. S. Brown and W. I. Weisberger, Nucl. Phys. {\bf B157}, 285 (1979).
 
\bibitem{strickland} S. Mrowczynski, A. Rebhan and M. Strickland, Phys. Rev. {\bf D70}, 025004 (2004).

\end{thebibliography}
\end{document}